\documentclass[twocolumn,superscriptaddress,nofootinbib,notitlepage,longbibliography,aps]{revtex4-1}
\pdfoutput=1

\usepackage{graphicx}
\usepackage{amsmath}
\usepackage{amssymb}
\usepackage{amsfonts}
\usepackage[dvipsnames]{xcolor}
\usepackage[linktoc=none]{hyperref}
\usepackage[utf8]{inputenc}
\usepackage{comment}

\definecolor{tab-blue}{RGB}{0, 107, 164}
\hypersetup{colorlinks=true,allcolors=tab-blue}

\newcommand{\INFN}{INFN - Sezione di Napoli, Complesso Universitario Monte S. Angelo, 80126 Napoli, Italy}
\newcommand{\SSM}{Scuola Superiore Meridionale, Via Mezzocannone 4, 80138 Napoli, Italy}

\begin{document}

\title{High-energy gamma-ray emission from memory-burdened primordial black holes}
\author{Marco Chianese}
\email{m.chianese@ssmeridionale.it}
\affiliation{\SSM}
\affiliation{\INFN}

\begin{abstract}
Theoretical studies on the memory-burden effect suggest that Primordial Black Holes (PBHs) with masses smaller than $10^{15}$ grams may be viable dark matter candidates and, consequently, be potential sources of high-energy particles in the present Universe. In this paper, we investigate the evaporation of memory-burdened PBHs into high-energy gamma-rays. Differently from previous analyses, we account for the attenuation of gamma-rays caused by their interaction with background radiation at energies above $10^5~{\rm GeV}$, as well as the secondary emission from the electromagnetic cascades generated during their propagation through extragalactic space. Performing a likelihood analysis with current gamma-ray data, we place new constraints on the parameter space of memory-burdened PBHs. Our results show that ultra-high-energy diffuse gamma-ray observations set more restrictive bounds than high-energy neutrino data, particularly in scenarios with a strong memory-burden suppression of the PBH evaporation.
\end{abstract}

\maketitle
\tableofcontents

\section{Introduction}

Primordial Black Holes (PBHs) are hypothetical black holes that may have formed in the early Universe through the direct collapse of primordial overdensities, preceding the formation of the first stars~\cite{Zeldovich:1967lct, Hawking:1971ei, Carr:1974nx}. Over the past decade, these mysterious objects have attracted growing interest, particularly in the search for Dark Matter (DM) and gravitational waves. An interesting aspect of PBHs is their instability caused by the emission of fundamental particles via Hawking radiation. Within the standard framework, only PBHs with masses exceeding $10^{15}$ grams would have lifetimes longer than the age of the Universe, making them viable DM candidates. Stringent constraints, however, restrict PBHs as the exclusive DM component of the Universe to the asteroid-mass range of $10^{17}~{\rm g}\lesssim M_{\rm PBH} \lesssim 10^{22}~{\rm g}$~\cite{Carr:2016drx, Green:2020jor, Carr:2020gox, Carr:2021bzv}. On the other hand, PBHs with $M_{\rm PBH} \lesssim 10^{15}~{\rm g}$ may still have phenomenological implications, as their existence could have significantly influenced several cosmological phenomena. For instance, PBHs could modify the generation of the baryon asymmetry of the Universe~\cite{Fujita:2014hha, Hamada:2016jnq, Morrison:2018xla, Chen:2019etb, Perez-Gonzalez:2020vnz, Datta:2020bht, Hooper:2020otu, JyotiDas:2021shi, DeLuca:2021oer, Bernal:2022pue, Calabrese:2023key, Calabrese:2023bxz, Schmitz:2023pfy, Barman:2024slw, Gunn:2024xaq}, produce gravitational waves~\cite{Papanikolaou:2020qtd, Domenech:2020ssp, Papanikolaou:2022chm, Ireland:2023avg, Domenech:2024wao}, and drive DM formation~\cite{Bernal:2020kse, Gondolo:2020uqv, Bernal:2020ili, Bernal:2020bjf, Cheek:2021odj, Cheek:2021cfe, Samanta:2021mdm, Bernal:2021yyb, Bernal:2021bbv, Sandick:2021gew, Bernal:2022oha, Cheek:2022mmy, Gehrman:2023qjn, Bertuzzo:2024fns}.

The semi-classical description of black hole evaporation assumes that the black hole remains classical throughout its lifetime~\cite{Hawking:1975vcx}. This model may lack self-consistency, as notably known in the context of the information loss paradox~\cite{Almheiri:2020cfm, Buoninfante:2021ijy}. Specifically, Hawking's original analysis neglects the influence of the emitted radiation on the quantum state of the black hole. However, this effect is expected to be significant when the energy of the emitted particles is comparable to the total energy of the black hole. Recently, Ref.s~\cite{Dvali:2018xpy, Dvali:2020wft, Dvali:2024hsb} have proposed that quantum back-reaction may significantly slow down the black hole evaporation, as universally observed in quantum systems where the retained information resists decay. This effect, known as ``memory burden'', would imply that PBHs with $M_{\rm PBH} \lesssim 10^{15}~{\rm g}$ could still be evaporating today, potentially leading to fascinating phenomenological consequences~\cite{Alexandre:2024nuo, Thoss:2024hsr, Haque:2024eyh, Balaji:2024hpu, Barman:2024iht, Bhaumik:2024qzd, Barman:2024ufm, Kohri:2024qpd, Jiang:2024aju, Chianese:2024rsn, Zantedeschi:2024ram, Barker:2024mpz, Borah:2024bcr, Loc:2024qbz, Basumatary:2024uwo, Athron:2024fcj, Bandyopadhyay:2025ast, Calabrese:2025sfh, Boccia:2025hpm, Liu:2025vpz, Dvali:2025ktz, Montefalcone:2025akm}.

In this paper, we thoroughly examine the diffuse emission of high-energy gamma-rays from the evaporation of memory-burdened PBHs, which may constitute a fraction of the DM component. Hence, we establish new limits on the parameter space of memory-burdened PBHs by analyzing the current upper bounds on the Ultra-High-Energy (UHE) diffuse gamma-ray flux from $10^5$ to $10^{11}$~GeV placed by several experimental collaborations, including CASA-MIA~\cite{CASA-MIA:1997tns}, KASCADE and KASCADE-Grande~\cite{KASCADEGrande:2017vwf}, Pierre Auger Observatory~\cite{PierreAuger:2015fol, PierreAuger:2016kuz}, and Telescope Array~\cite{TelescopeArray:2018rbt}. Moreover, we revisit previous constraints~\cite{Thoss:2024hsr, Haque:2024eyh} from Fermi-LAT~\cite{Fermi-LAT:2014ryh} and LHAASO~\cite{LHAASO:2023gne} measurements below $10^6$~GeV by taking into account the gamma-ray attenuation and the secondary emission from the electromagnetic cascades. At energies higher than $\sim 10^{5}~{\rm GeV}$, the interaction of gamma-rays with galactic and extragalactic background radiation becomes kinematically possible, resulting in the creation of electron-positron pairs. This process prevents a fraction of gamma-rays from reaching the Earth, thus attenuating the gamma-ray flux from evaporating PBHs. Furthermore, electrons and positrons undergo interactions such as Inverse Compton scattering with background photons, bremsstrahlung, and synchrotron radiation in the galactic and intergalactic magnetic fields. These interactions, which occur multiple times during propagation, generate electromagnetic cascades that eventually produce a secondary flux of low-energy photons. We demonstrate here that this component represents the dominant contribution in the Fermi-LAT energy range, as the primary gamma-ray emission is strongly suppressed by attenuation.

The paper is organized as follows. In Sec.~\ref{sec:memory} we describe the Hawking evaporation of memory-burdened PBHs. In Sec.~\ref{sec:gsamma} we discuss in detail the computation of the primary and secondary gamma-ray flux from galactic and extragalactic distributions of PBHs as DM component. In Sec.~\ref{sec:results} we report the constraints placed with current gamma-ray data. Finally, in Sec.~\ref{sec:conclusions} we draw our conclusions.

\section{Memory-burdened primordial black holes} \label{sec:memory}
\begin{figure*}[t!]
    \centering
    \includegraphics[width=0.75\linewidth]{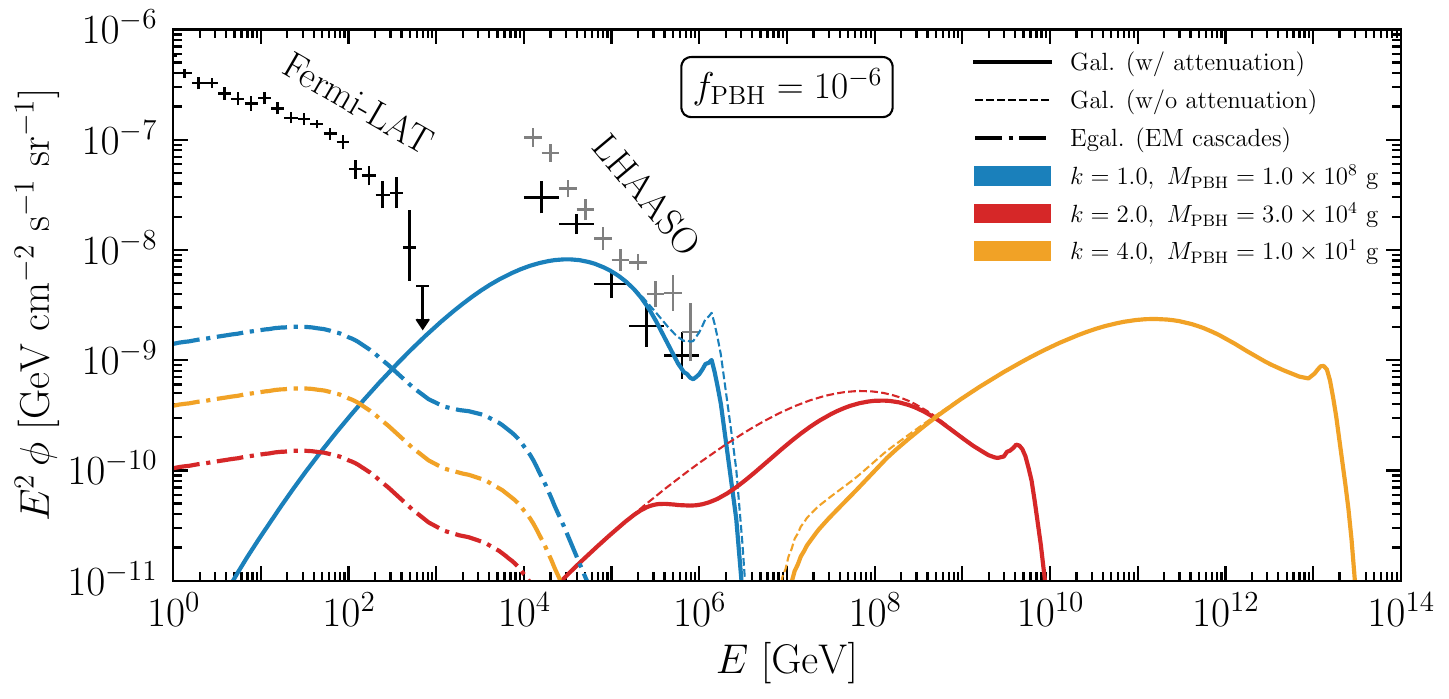}
    \caption{{\bf Diffuse gamma-ray flux.} The different colors correspond to different combinations of the initial PBH mass $M_{\rm PBH}$ and the memory-burden parameter $k$, taking $f_{\rm PBH} = 10^{-6}$. The solid/dashed lines show the prompt galactic gamma-ray flux (see Eq.~\eqref{eq:gal}) with/without the effect of gamma-ray attenuation, while the dot-dashed lines show the universal spectrum of the secondary emission from extragalactic electromagnetic cascades (see Eq.~\eqref{eq:egal}). The fluxes are averaged over $\Delta\Omega=4\pi$. From left to right, the data points correspond to the isotropic gamma-ray background measured by Fermi-LAT~\cite{Fermi-LAT:2014ryh} and the galactic emission observed by LHAASO~\cite{LHAASO:2023gne} (inner and outer regions in gray and black colors, respectively).}
    \label{fig:benchmark_diffuse}
\end{figure*}

It is widely recognized that black holes undergo particle emission due to quantum effects. This radiation follows an approximately thermal spectrum determined by the Hawking temperature:
\begin{equation}
    T_{\rm H} = \frac{1}{8 \pi G M_{\rm PBH}} \simeq 10^{4}\left(\frac{10^9~{\rm g}}{M_{\rm PBH}}\right)~{\rm GeV}\,,
    \label{eq:haw}
\end{equation}
where $G$ denotes the gravitational constant. The energy fueling this emission originates from the PBH gravitational field according to a mass-loss rate given by
\begin{equation}
    \frac{{\rm d}M_{\rm PBH}}{{\rm d}t} = - \frac{\mathcal{G} \,g_{\rm SM} }{30720 \pi \, G^2 M_{\rm PBH}^2}\,.
    \label{eq:mbrate}
\end{equation}
Here, $\mathcal{G} \approx 3.8$ accounts for gray-body effects due to gravitational back-scattering~\cite{Page:1976wx}, while $g_{\rm SM} \approx 102.6$ represents the number of relativistic degrees of freedom at the temperature $T_{\rm H}$ in the Standard Model~\cite{Mazde:2022sdx}. In the conventional scenario, the evaporation process continues until the entire PBH mass is transformed into radiation. The time required for complete evaporation results to be
\begin{equation}
\tau_{\rm PBH} = \frac{10240 \pi G^2 M_{\rm PBH}^3}{\mathcal{G} g_H} \simeq 4.4\times 10^{17}\left(\frac{M_{\rm PBH}}{10^{15}~{\rm g}}\right)^3~{\rm s}\,.
\end{equation}
This implies that PBHs with masses below $10^{15}$ g would have fully evaporated within the age of the Universe.

When memory burden is taken into account, PBH evaporation instead progresses in two distinct phases. The first is a semi-classical Hawking-like stage, wherein the PBH follows standard mass loss dynamics. The second is a ``burdened phase,'' during which quantum memory effects decelerate the evaporation process. In the instantaneous case, the transition between these phases occurs at a time
\begin{equation}
t_q = \tau_{\rm PBH}(1-q^3)\,,
\end{equation}
leaving the PBH with a fraction $q$ of its initial mass, \textit{i.e.}
\begin{equation}
M^{\rm mb}_{\rm PBH} = q\,M_{\rm PBH}\,.
\end{equation}
For times $t \geq t_q$, the stored information on the PBH event horizon induces a back-reaction that reduces the mass-loss rate according to
\begin{equation}
\frac{{\rm d}M^{\rm mb}_{\rm PBH}}{{\rm d}t} = \frac{1}{S(M_{\rm PBH})^{k}}\frac{{\rm d}M_{\rm PBH}}{{\rm d}t}\,,
\end{equation}
with $k>0$ and the quantity $S$ being the PBH entropy defined as
\begin{equation}
S(M_{\rm PBH}) = 4 \pi G M_{\rm PBH}^2\,.
\end{equation}
Solving this equation gives the PBH mass evolution during the memory-burdened phase:
\begin{equation}
M_{\rm PBH}^{\rm mb}(t) = M_{\rm PBH}^{\rm mb} \left[ 1- \Gamma_{\rm PBH}^{(k)}(t-t_q)\right]^{1/(3+2k)}\,,
\end{equation}
where
\begin{equation}
\Gamma_{\rm PBH}^{(k)} = \frac{\mathcal{G}\,g_{\rm SM}}{7680 \pi}2^k(3+2k)M_P\left(\frac{M_{\rm P}}{M_{\rm PBH}^{\rm mb}}\right)^{3+2k}\,,
\end{equation}
with $M_{\rm P}=1/\sqrt{8 \pi G}$ being the reduced Planck mass. Consequently, the total evaporation time extends to
\begin{equation}
\tau_{\rm PBH}^{(k)} = t_q + (\Gamma_{\rm PBH}^{(k)})^{-1} \simeq (\Gamma_{\rm PBH}^{(k)})^{-1}\,.
\end{equation}
This timescale can be significantly longer than in the standard scenario, allowing much lighter PBHs to persist to the present day. As a result, these objects could contribute to the current DM density.

For this analysis, we assume $q = 1/2$ as the memory-burden effect is expected to become significant when half of the initial PBH mass has been lost. It is worth noting that the precise choice of $q$ only influences how our results are interpreted in terms of the original PBH mass. Moreover, we assume an instantaneous transition from the semi-classical regime to the memory-burdened one. Recent studies~\cite{Dvali:2025ktz, Montefalcone:2025akm} have demonstrated that a non-instantaneous transition could significantly alter the PBH evolution throughout the entire history of the Universe. However, as will be discussed in the following sections, the gamma-ray constraints are primarily based on the current galactic emission. Consequently, the limits we report in this paper can be rescaled to account for the specific suppression of PBH evaporation in different non-instantaneous models. A dedicated analysis will be provided in future work.

\section{The high-energy gamma-ray emission} \label{sec:gsamma}

We analyze the flux of high-energy gamma-rays produced by a population of memory-burdened PBHs, which contribute a fraction $f_{\rm PBH}=\Omega_{\rm PBH}/\Omega_{\rm DM}$ to the total DM density of the today's Universe. We consider a monochromatic PBH mass spectrum within the range $10^{-1}~{\rm g} \leq M_{\rm PBH} \leq 10^9 ~{\rm g}$.

The semi-classical gamma-ray emission rate from a non-rotating, neutral PBH of mass $M_{\rm PBH}$ is given by  
\begin{equation}
    \frac{{\rm d}^2N_{\gamma}}{{\rm d}E{\rm d}t} = \frac{g_{\gamma}}{2 \pi} \frac{\mathcal{F}(E,M_{\rm PBH})}{e^{E/T_{\rm H}} - 1} \,,
\end{equation}
where $g_{\gamma} = 2$ accounts for the internal degrees of freedom of the photons, and $\mathcal{F}(E,M_{\rm PBH})$ is the gray-body factor. We numerically compute this emission rate using the code \texttt{BlackHawk}~\cite{Arbey:2019mbc, Arbey:2021mbl}, which also incorporates secondary photon production via the code \texttt{HDMSpectra}~\cite{Bauer:2020jay}.

During the memory-burdened phase, the emission rate is suppressed as described by Eq.~\eqref{eq:mbrate}, implying
\begin{equation}
    \frac{{\rm d}^2N^{\rm mb}_{\gamma}}{{\rm d}E{\rm d}t}  = \frac{1}{S(M_{\rm PBH})^{k}} \, \frac{{\rm d}^2N_{\gamma}}{{\rm d}E{\rm d}t} \,.
    \label{eq:mb_spectrum}
\end{equation}
Although the overall emission is reduced, the spectral peak remains at the temperature of the order of $T_{H}$, as in the semi-classical regime. As a result, photons emitted today from surviving memory-burdened PBHs with $M_{\rm PBH} \lesssim 10^{9}~{\rm g}$ would possess energies of at least $E \gtrsim 10~{\rm TeV}$.

At these high energies, gamma-rays suffer from large attenuation due to the interactions with background radiation. Therefore, the total gamma-ray flux from memory-burdened PBHs consists of two main contributions: \textit{i}) the prompt attenuated emission from the galactic DM halo, and \textit{ii}) the secondary emission due to the electromagnetic cascades initiated by primary photons from the extragalactic DM distribution. While the former dominates at very high energies $(E \sim \mathcal{O}(T_{\rm H}))$, the latter is relevant for $E\lesssim 10^{5}~{\rm GeV}$. We have checked that the prompt emission from the extragalactic DM distribution is highly negligible due to the gamma-ray attenuation.

The galactic prompt component reads
\begin{equation}
    \frac{{\rm d}^2\phi^{\rm gal}_{\gamma}}{{\rm d}E{\rm d}\Omega} = \frac{f_{\rm PBH}}{4 \pi M^{\rm mb}_{\rm PBH}} \frac{{\rm d}^2 N^{\rm mb}_{\gamma}}{{\rm d}E{\rm d}t} \mathcal{J}\left(E_\gamma,\Delta \Omega\right)\,,
    \label{eq:gal}
\end{equation}
where the normalization of the flux and the gamma-ray emission rate are defined by the today PBH mass $M_{\rm PBH}^{\rm mb}$, and $\mathcal{J}$ is the energy-dependent J-factor accounting for the gamma-ray attenuation. Averaging over a solid angle $\Delta \Omega$, we have
\begin{equation}
    \mathcal{J} = \frac{1}{\Delta \Omega}\int_{\Delta \Omega} {\rm d}\Omega \int_0^\infty {\rm d}s ~\rho_\mathrm{DM}(r)e^{-\tau_{\gamma\gamma}(E_\gamma,s,b,l)} \,,
    \label{eq:Jfactor}
\end{equation}
where $\rho_\mathrm{DM}(r)$ is the galactic DM halo density profile as a function of the galactocentric radial coordinate $r= (s^2+R_\odot^2-2sR_\odot\cos b\cos l)^{1/2}$ with $R_\odot= 8.178~{\rm kpc}$ being the Sun distance from the galactic center, and $\tau_{\gamma\gamma}$ is the optical depth due to the production of electron-positron pairs via the interaction with background photons. We consider a NFW density profile
\begin{equation}
    \rho_\mathrm{DM}(r) = \frac{\rho_s}{r/r_s(1+r/r_s)^2}\,,
\end{equation}
with scale radius $r_s = 25~{\rm kpc}$ and $\rho_s=0.23 ~{\rm GeV/cm^3}$ providing a local DM density of $0.4~{\rm GeV/cm^3}$~\cite{Iocco:2015xga, Benito:2019ngh, Benito:2020lgu}. This choice also ensures a consistent comparison with the constraints recently placed using high-energy neutrinos~\cite{Chianese:2024rsn}. The optical depth in Eq.~\eqref{eq:Jfactor} is computed following Ref.s~\cite{Capanema:2021hnm, Chianese:2021jke}. We take into into account both the homogeneous cosmic microwave background (CMB) and the galactic starlight and infrared radiation as taken from the \texttt{GALPROPv54} code~\cite{Porter:2017vaa}. Although the galactic radiation varies across the Milky Way, and peaks toward the galactic center and along the galactic plane, the angular dependence of the J-factor is primarily determined by the DM density profile. We find that, for $\Delta\Omega=4\pi$, the averaged J-factor at $10^6~{\rm GeV}$ is $\mathcal{J} = 7.84 \times 10^{21}~{\rm GeV/cm^2}$ and $\mathcal{J} = 2.22 \times 10^{22}~{\rm GeV/cm^2}$ with and without the gamma-ray attenuation, respectively.

The secondary extragalactic emission can be defined as
\begin{equation}
    \frac{{\rm d}^2\phi^{\rm egal}_{\gamma}}{{\rm d}E{\rm d}\Omega} = \frac{f_{\rm PBH}\,\Omega_{\rm DM}\,\rho_{\rm c}}{M^{\rm mb}_{\rm PBH}} \left.\frac{{\rm d}^2\phi^{\rm EM}_{\gamma}}{{\rm d}E{\rm d}\Omega}\right|^{z_{\rm max}}_{z=0}\,,
    \label{eq:egal}
\end{equation}
where the first term fixes the normalization with $\Omega_{\rm DM} = 0.264$ and $\rho_{\rm c} = 4.79\times 10^{-6}~{\rm GeV/cm^3}$~\cite{Planck:2018vyg}, and the last term corresponds to an isotropic spectrum with a nearly universal shape, which depends on the CMB and the extragalactic background light (EBL)~\cite{Berezinsky:2016feh}. We numerically compute this spectrum by means of the \texttt{$\gamma$-CascadeV4} code~\cite{Blanco:2018bbf, Capanema:2024nwe}, providing as input the injected gamma-ray spectrum given in Eq.~\eqref{eq:mb_spectrum} and assuming the best-fit EBL model from Ref.~\cite{Saldana-Lopez:2020qzx}. We take into account the time evolution $M_{\rm PBH}(t)$ of the PBH mass and integrate up to a redshift $z_{\rm max}=10$, which is the maximum allowed value by the \texttt{$\gamma$-CascadeV4} code. By varying $z_{\rm max}$ from 1 to 10, we have verified that the secondary extragalactic emission remains unchanged at $E > 10^2~{\rm GeV}$. This indicates that contributions from higher redshifts would have no impact on the Fermi-LAT limits which are mainly driven by the flux at these energies.

In Fig.~\ref{fig:benchmark_diffuse} we show the prompt galactic emission defined in Eq.~\eqref{eq:gal} (solid lines) and the secondary extragalactic emission defined in Eq.~\eqref{eq:egal} (dot-dashed lines) in case of three different choices for the initial PBH mass and the parameter $k$, assuming $f_{\rm PBH} = 10^{-6}$. In order to highlight the effect of gamma-ray attenuation, which is relevant from $10^5$ to $10^9$~GeV, we also show the prompt galactic emission without attenuation with thin dashed lines.
The data points correspond to the isotropic gamma-ray background (IGRB) measured by Fermi-LAT~\cite{Fermi-LAT:2014ryh} (black points below $10^3$~GeV) and to the diffuse emission from an inner region ($15^\circ < l < 125^\circ$ and $|b|< 5^\circ$, gray points) and an outer region ($125^\circ < l < 235^\circ$ and $|b|< 5^\circ$, black points) of the Milky Way recently measured by LHAASO~\cite{LHAASO:2023gne}.
We find that the smaller the PBH mass, the higher the energies of the gamma-ray emitted by the galactic PBH distribution, which is typically well beyond the Fermi-LAT energy range. This implies that the Fermi-LAT and LHAASO telescopes can mainly probe the secondary extragalactic emission and the prompt galactic emission, respectively.
\begin{figure}[t!]
    \centering
    \includegraphics[width=\linewidth]{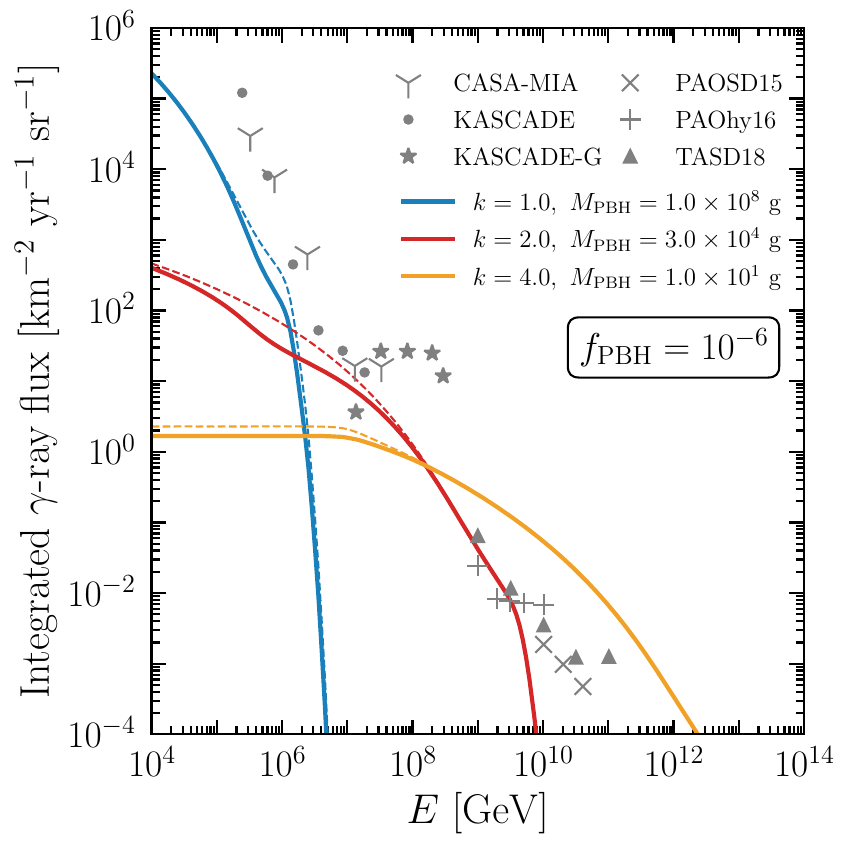}
    \caption{{\bf Integrated gamma-ray flux.} The different colors correspond to different combinations of the initial PBH mass $M_{\rm PBH}$ and the memory-burden parameter $k$, taking $f_{\rm PBH} = 10^{-6}$. These fluxes are obtained from Eq.~\eqref{eq:integrated} with/without taking into account the gamma-ray attenuation (solid/dashed lines). The fluxes are averaged over $\Delta\Omega=4\pi$. Gray data points are the upper limits on the UHE gamma-ray flux placed by CASA-MIA~\cite{CASA-MIA:1997tns}, KASCADE and KASCADE-Grande~\cite{KASCADEGrande:2017vwf} at 90\% CL, and by Pierre Auger Observatory (PAO)~\cite{PierreAuger:2015fol, PierreAuger:2016kuz} and Telescope Array Surface Detector (TASD)~\cite{TelescopeArray:2018rbt} at 95\% CL.}
    \label{fig:benchmark_integrated}
\end{figure}

In Fig.~\ref{fig:benchmark_integrated} we report the integrated gamma-ray flux 
\begin{equation}
    \Phi_\gamma = \int_{E}^\infty {\rm d}E^\prime~\frac{{\rm d}^2\phi^{\rm gal}_{\gamma}}{{\rm d}E^\prime{\rm d}\Omega}\,,
    \label{eq:integrated}
\end{equation}
in case of the three benchmark scenarios previously discussed. In this energy range, the gamma-ray emission is dominated by the galactic component, while the extragalactic contribution is highly negligible. As before, the solid and dashed lines represent the flux with and without the gamma-ray attenuation, respectively. In the figure, we also report a collection of upper bounds placed by several experiments~\cite{CASA-MIA:1997tns, KASCADEGrande:2017vwf, PierreAuger:2015fol, PierreAuger:2016kuz, TelescopeArray:2018rbt}.

\section{Results}\label{sec:results}
\begin{figure*}[t!]
    \centering
    \includegraphics[width=\linewidth]{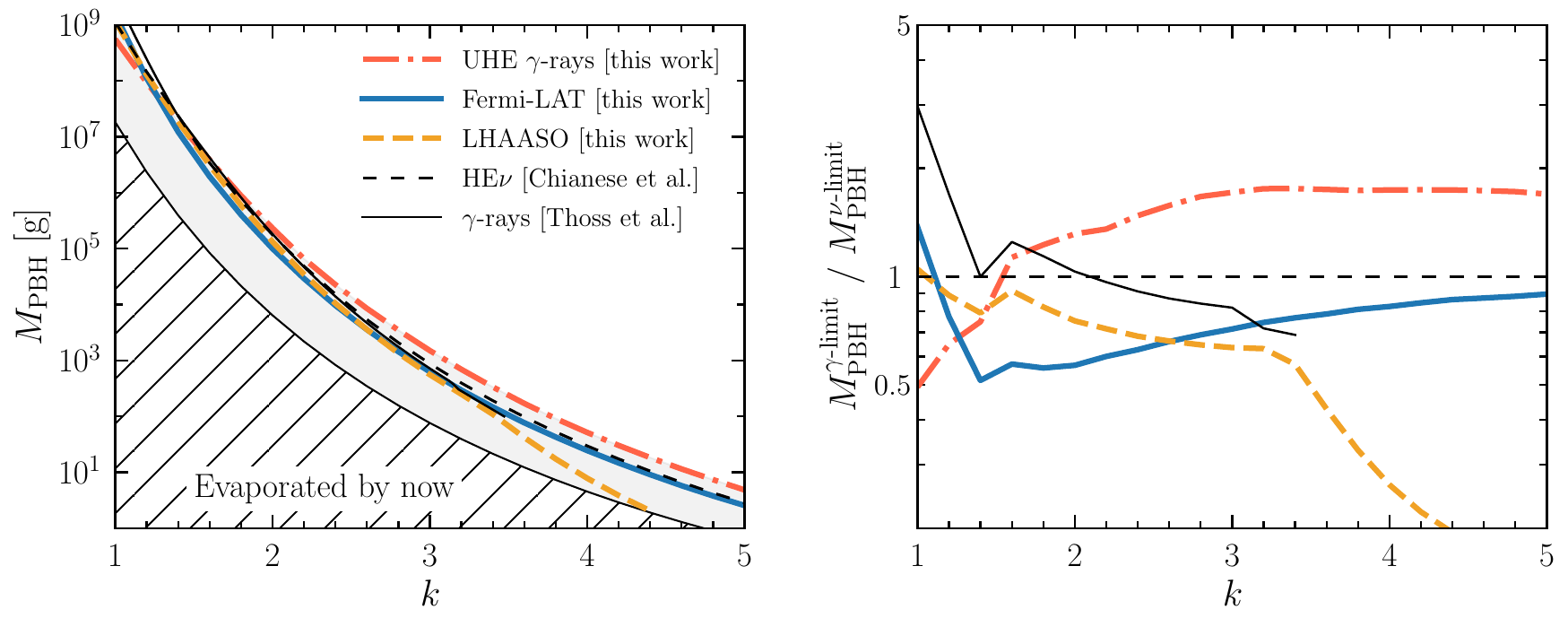}
    \caption{{\bf Gamma-ray constraints on memory-burdened PBHs as viable DM candidates.} \textit{Left:} the colored lines refer to gamma-ray limits placed at 95\% C.L. in the $M_{\rm PBH}$-$k$ plane with $f_{\rm PBH} = 1$, in case of Fermi-LAT~\cite{Fermi-LAT:2014ryh} (solid blue line), LHAASO~\cite{LHAASO:2023gne} (dashed yellow line) and UHE diffuse gamma-ray~\cite{CASA-MIA:1997tns, KASCADEGrande:2017vwf, PierreAuger:2015fol, PierreAuger:2016kuz, TelescopeArray:2018rbt} (dot-dashed red line) data. The white area represents memory-burdened PBHs that can serve as viable DM candidates ($f_{\rm PBH} = 1$), whereas the hatched region indicates PBHs that have fully evaporated in cosmological times. Previous gamma-ray~\cite{Thoss:2024hsr} and neutrino limits~\cite{Chianese:2024rsn} are shown by the thin black lines with solid and dashed styles, respectively. \textit{Right:} ratio of the gamma-ray limits on $M_{\rm PBH}$ reported in the left panel to the neutrino constraint previously obtained in Ref.~\cite{Chianese:2024rsn}, which is displayed as reference with the horizontal dashed line.}
    \label{fig:limits}
\end{figure*}\begin{figure*}[t!]
    \centering
    \includegraphics[width=\linewidth]{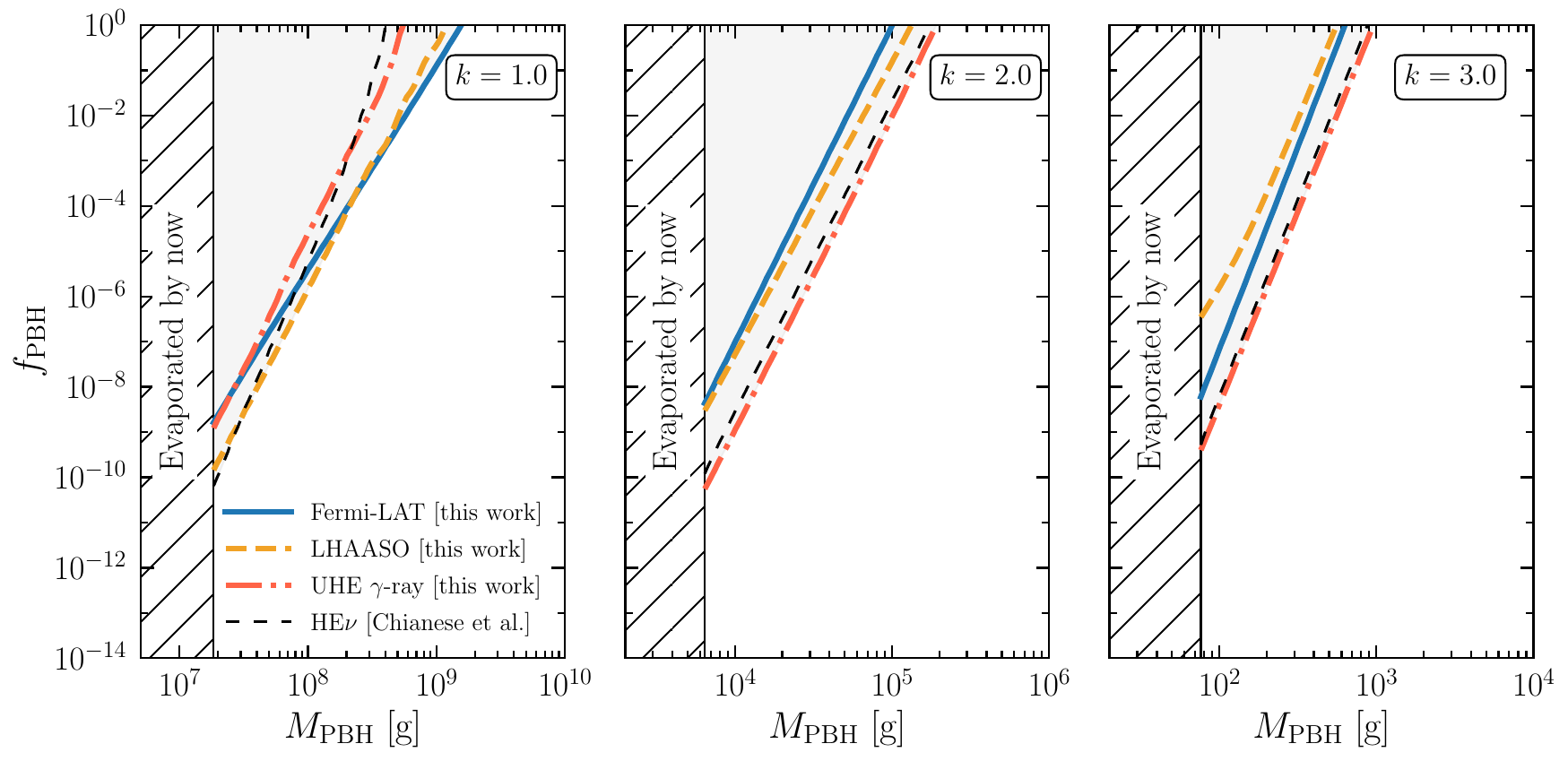}
    \caption{{\bf Gamma-ray constraints on the PBH abundance.} Constraints at 95\% CL in the $M_{\rm PBH}$-$f_{\rm PBH}$ plane for different values of the memory-burden parameter $k$ ($k=1.0$ left panel, $k=2.0$ middle panel, and $k=3.0$ right panel). Lines with different shading refer to different data samples: Fermi-LAT~\cite{Fermi-LAT:2014ryh} (solid blue line), LHAASO~\cite{LHAASO:2023gne} (dashed yellow line), UHE diffuse gamma-ray~\cite{CASA-MIA:1997tns, KASCADEGrande:2017vwf, PierreAuger:2015fol, PierreAuger:2016kuz, TelescopeArray:2018rbt} (dot-dashed red line), and high-energy neutrinos~\cite{Chianese:2024rsn}. The hatched regions indicates PBHs that have fully evaporated in cosmological times.}
    \label{fig:limits_k}
\end{figure*}

We analyze the data reported by each experimental collaboration reported in Fig.s~\ref{fig:benchmark_diffuse} and~\ref{fig:benchmark_integrated} to place constraints on the PBH abundance $f_{\rm PBH}$ as a function of the parameters $\vec{\theta}_{\rm PBH}=(M_{\rm PBH}, k)$. We assume that the data are well explained by the gamma-ray emission from astrophysical sources, while the PBH evaporation provides an additional contribution (see e.g. Ref.s~\cite{Ajello:2015mfa,Ambrosone:2020evo,Roth:2021lvk} and~\cite{Lipari:2018gzn,DeLaTorreLuque:2025zsv} for the astrophysical interpretation of the Fermi-LAT and LHAASO observations, respectively]).
For each experiment, we therefore consider the following background-agnostic likelihood function 
\begin{equation}
    \mathcal{L}\left(f_{\rm PBH}; \vec{\theta}_{\rm PBH} \right) = \prod_{i}^{n_{\rm data}}\left\{\begin{array}{l l}
    \mathcal{P}\left(d_i|\mu_i, \sigma_i\right) & \mu_i > d_i \\
    1 &\mu_i \leq d_i 
    \end{array}\right. \,.
\end{equation}
The probability distribution function $\mathcal{P}$ of the flux data $d_i$ is assumed to be a Gaussian distribution with expected mean $\mu_i(f_{\rm PBH}; \vec{\theta}_{\rm PBH})$ and standard deviations $\sigma_i$ defined by the experimental data. Depending on the data sample, the expected mean is either the diffuse gamma-ray emission from galactic and extragalactic DM distributions given in Eq.s~\eqref{eq:gal} and~\eqref{eq:egal}, respectively, or the integrated gamma-ray flux given in Eq.~\eqref{eq:integrated}. We also take into account the different sky coverage of each experiment by suitably choosing the solid angle $\Delta\Omega$ in Eq.~\eqref{eq:Jfactor}. When reported, we also include the systematic uncertainties in the determination of the quantities $\sigma_i$. Thus, for each selected combination of $M_{\rm PBH}$ and $k$, we determine the maximum allowed value of $f_{\rm PBH}$ by evaluating  $\Delta \chi^2 = -2\ln \mathcal{L}$ and applying Wilks' theorem under the assumption of a single degree of freedom.

In Fig.s~\ref{fig:limits} and~\ref{fig:limits_k}, we report the main results of the present analysis. In the left panel of Fig.~\ref{fig:limits}, the different lines illustrate the constraints on the PBH mass $M_{\rm PBH}$ as a function of the memory-burden parameter $k$, assuming $f_{\rm PBH} = 1$. This implies that the white region above the lines corresponds to the parameter space where PBHs can fully constitute the DM component of the Universe. The hatched region represents the parameter space where PBHs have completely evaporated over cosmological timescales. The colored lines depict the constraints we have placed employing current gamma-ray data with energies above GeV, while the black thin dashed and solid lines show the previous constraints from high-energy neutrinos (HE$\nu$)~\cite{Chianese:2024rsn} and gamma-rays~\cite{Thoss:2024hsr}, respectively. The HE$\nu$ constraints have been recomputed to correct  an issue in the implementation of the \texttt{HDMSpectra} code within the \texttt{BlackHawk} package, as identified by Ref.~\cite{Dondarini:2025ktz}.
In the right panel of the same figure, we show the ratio of the different constraints on $M_{\rm PBH}$ to the previous neutrino limit~\cite{Chianese:2024rsn}. In Fig.~\ref{fig:limits_k}, we show the constraints on the PBH abundance as a function of the PBH mass for three different memory-burden scenarios with the parameter $k$ fixed to 1.0 (left panel), 2.0 (middle panel) and 3.0 (right panel). The line styles are the same as in the previous figure.

We highlight a few key aspects of our results. First, we find that the constraints we place with Fermi-LAT and LHAASO data significantly differ from the ones obtained in previous analyses. For $k \lesssim 3.0$, they are weaker than the bounds placed by Ref.~\cite{Thoss:2024hsr}, while for $k \gtrsim 3.0$ the Fermi-LAT limit results to be more stringent. This is due to the fact that the gamma-ray attenuation and the secondary emission from extragalactic electromagnetic cascades were not previously considered. We also note that the LHAASO telescope probes the scenario of memory-burden PBHs up to $k\simeq 4.4$. Higher values for the parameter $k$ correspond to a prompt galactic gamma-ray flux above $10^6~{\rm GeV}$, which is outside the LHAASO energy range.
Second, the new limits imposed by the UHE gamma-ray data provide the most stringent constraints on the PBH parameter space, particularly for large values of the parameter $k$.
Lastly, our results enable a consistent comparison between gamma-ray and neutrino constraints, as they are both derived under the same assumptions, \textit{e.g.} for the DM distribution and the likelihood analysis. We robustly demonstrate that the current neutrino observations provide stronger constraints than gamma-ray data for $1.2 \lesssim k\lesssim 1.5$ only.

\section{Conclusions} \label{sec:conclusions}

We have explored the high-energy gamma-ray emission produced by the evaporation of memory-burdened Primordial Black Holes (PBHs). Thanks to their much longer lifetime, memory-burdened PBHs with a mass smaller than $10^{15}~{\rm g}$ could have survived to the present era, potentially becoming an important source of high-energy particles in the current Universe. By extending previous research, we have incorporated the impact of gamma-ray attenuation at energies exceeding $10^{5}~{\rm GeV}$, which weakens the gamma-ray constraints. Additionally, we have considered the contribution of the secondary radiation produced by electromagnetic cascades, which is relevant at energies smaller than $10^{5}~{\rm GeV}$. 

Using up-to-date high-energy diffuse gamma-ray data, we have derived new and tighter constraints on the scenario of memory-burdened PBHs. Our findings reveal that the ultra-high-energy diffuse gamma-ray observations impose the most stringent limits for a large portion of the parameter space of memory-burdened PBHs. These constraints provide highly competitive and complementary probes of the memory-burden effect, which are essential for advancing our understanding of the potential role of PBHs in the present Universe.

\section*{Data availability}

The gamma-ray and neutrino bounds presented in this article are publicly available at~\cite{Paperdata}.

\section*{Acknowledgements}

We thank Antonio Capanema for assistance with the \texttt{$\gamma$-Cascade} code.
We acknowledge the support by the research project TAsP (Theoretical Astroparticle Physics) funded by the Istituto Nazionale di Fisica Nucleare (INFN).

\bibliography{bibliography}
\end{document}